\input harvmac
\noblackbox
\def\Title#1#2{\rightline{#1}\ifx\answ\bigans\nopagenumbers\pageno0\vskip1in
\else\pageno1\vskip.8in\fi \centerline{\titlefont #2}\vskip .5in}

\font\bbbi=msbm10
\def\bbb#1{\hbox{\bbbi #1}}
\def\msurr{\mathsurround=0pt}
\def\overleftrightarrow#1{\vbox{\msurr\ialign{##\crcr
	$\leftrightarrow$\crcr\noalign{\kern-1pt\nointerlineskip}
	$\hfil\displaystyle{#1}\hfil$\crcr}}}

\def\ads{anti-de Sitter}
\def\mads{$AdS_{d+1}$}
\def\gbdel{{G_{B\partial}}}
\def\kbdel{{K_{B\partial}}}
\def\sq{{\vbox {\hrule height 0.6pt\hbox{\vrule width 0.6pt\hskip 3pt
   \vbox{\vskip 6pt}\hskip 3pt \vrule width 0.6pt}\hrule height 0.6pt}}}
\def\vecm{{\vec m}}
\def\vecx{{\vec x}}
\def\veck{{\vec k}}
\def\hf{{1\over 2}}

\def\limrho{\buildrel \rho\rightarrow \pi/2 \over \longrightarrow}

\def\limR{\buildrel  \longrightarrow \over {\scriptscriptstyle R\rightarrow \infty}}

\def\calo{{\cal O}}
\def\caln{{\cal N}}

\def\khat{{\hat K}}
\def\ehat{{\hat e}}
\def\onl{{\omega_{nl}}}
\def\Anl{{A_{nl}}}
\def\nlm{{nl\vecm}}
%
%journals
%
\def\ajou#1&#2(#3){\ \sl#1\bf#2\rm(19#3)}
\def\jou#1&#2(#3){,\ \sl#1\bf#2\rm(19#3)}
%%%%%%%%%
%
%fonts
%
\font\bbbi=msbm10
\font\ticp=cmcsc10
%%%%%%%%%
%
%References
%
\lref\GKP{S.S. Gubser, I. Klebanov and A. Polyakov, ``Gauge
theory correlators from noncritical string theory'', hep-th/9802109,
\ajou Phys. Lett. &B428 (98) 105.}

\lref\Mart{A. Martin, ``Minimal interactions at very high transfers,''
\ajou Nuov. Cim. &37 (65) 671.}

\lref\Mald{J. Maldacena, ``The large-N limit of superconformal
field theories and supergravity'', hep-th/9711200, \ajou
Adv. Theor. Math. Phys. &2 (98) 231.}

\lref\Witt{E. Witten, ``Anti-de Sitter space and holography'',
hep-th/9802150, \ajou Adv. Theor. Math. Phys. &2 (98) 253.}

\lref\BKL{V. Balasubramanian, P. Kraus and A. Lawrence,
``Bulk vs. Boundary dynamics in anti-de Sitter spacetime'',
hep-th/9805171, \ajou Phys. Rev & D59:046003 (99).}

\lref\BKLT{V. Balasubramanian, P. Kraus, A. Lawrence and S. Trivedi,
``Holographic probes of anti-de Sitter space-times'',
hep-th/9808017, {\sl Phys. Rev.} {\bf D59:104021} (1999).}

\lref\FMMR{D.Z. Freedman, S.D. Mathur, A. Matusis, and L. Rastelli, 
 ``Correlation functions in the CFT(d)/AdS(d+1) correspondence,'' 
hep-th/9804058,\ajou Nucl. Phys. &B546 (99) 96.}

\lref\InOo{T. Inami and H. Ooguri, ``One loop effective potential
 in anti-de Sitter space,''\ajou Prog. Theor. Phys. &73 (85) 1051.}

\lref\BuLu{C.P. Burgess and C.A. Lutken, ``Propagators and effective
 potentials in anti-de Sitter
space,''\ajou Phys. Lett. &153B (85) 137.}

\lref\Li{M. Li, ``Energy momentum conservation and holographic S matrix,''
hep-th/9904164.}

\lref\Froi{M. Froissart, Asymptotic behavior 
and subtractions in the Mandelstam
representation,''\ajou Phys. Rev. &123 (61) 1053.}

\lref\GrMe{D.J. Gross and P.F. Mende, ``The high-energy behavior 
of string scattering amplitudes,''\ajou Phys. Lett. &197B (87) 129; 
``String theory beyond the Planck scale,''\ajou Nucl.
Phys. &B303 (88) 407.}

\lref\Amat{D. Amati, M. Ciafaloni, and G. Veneziano, ``Superstring 
collisions at planckian energies,''\ajou Phys. Lett. &197B (87) 81.}

\lref\MeOo{P.F. Mende and H. Ooguri, ``Borel summation of string theory for 
Planck scale scattering,''\ajou  Nucl. Phys. &B339 (90) 641.}

\lref\FaSo{V. Ya. Fainberg and M.A. Solovev, ``How can local 
properties be described in field theories without strict
locality,''\ajou Ann. Phys. &113 (78) 421.}

\lref\CeMa{F. Cerulus and A. Martin, ``A lower bound for large angle elastic 
scattering at high energies,'' \ajou Phys. Lett. &8 (64) 80.}

\lref\AGMOO{O. Aharony,  S. Gubser, J. Maldacena, H. Ooguri,
and Y. Oz, ``Large N field theories, string theory and gravity,''
hep-th/9905111, to appear in {\sl Phys. Rep.}}

\lref\tHoo{G. 't Hooft, ``Dimensional reduction in
quantum gravity'', gr-qc/9310026.}

\lref\BDHM{T. Banks, M. Douglas, G. Horowitz and E. Martinec,
``AdS dynamics from conformal field theory'', hep-th/9808016,
UCSB/ITP preprint NSF-ITP-98082, U. Chicago preprint EFI-98-30.} 

\lref\Polc{J. Polchinski, ``S-matrices from AdS spacetime'',
hep-th/9901076.}

\lref\SussS{L. Susskind, ``Holography in the flat-space 
limit'', hep-th/9901079.}

\lref\DuFr{D.W. D\"usedau and D.Z. Freedman,
``Lehmann spectral representation for anti-de
Sitter quantum field theory'', \ajou Phys. Rev. &D33 (86) 389.}

\lref\Araf{I.Ya. Aref'eva, ``On the holographic S-matrix,'' hep-th/9902106.}

\lref\PoSu{J. Polchinski and L. Susskind, ``Puzzles and paradoxes about
holography,'' hep-th/9902182.}

\lref\HoIt{G. Horowitz and N. Itzhaki, ``Black holes, shock waves, and
causality in the AdS/CFT correspondence,'' hep-th/9901012.}

\lref\BGL{V. Balasubramanian, S.B. Giddings, and A. Lawrence, ``What do
CFTs tell us about anti-de Sitter spacetimes?'' hep-th/9902052, {\sl
JHEP} {\bf 9903:001} (1999).}

\lref\bsm{S.B. Giddings, ``The boundary S-matrix and 
the AdS to CFT dictionary,''
hep-th/9903048.}

\lref\AbSt{M. Abramowitz and I.A. Stegun,
{\sl Handbook of Mathematical Functions}, Dover (NY) 1965.}

\lref\Suss{L. Susskind, ``The world as a hologram'', hep-th/9409089,
{\sl J. Math. Phys.} {\bf 36} (1995) 6377.}

\Title{\vbox{\baselineskip12pt\hbox{hep-th/9907129}
%\hbox{Draft -- do not distribute}
}}
{\vbox{\centerline {Flat-space scattering and bulk locality}
\vskip2pt\centerline{ in the the AdS/CFT correspondence}
}}
\centerline{{\ticp Steven B. Giddings}\footnote{$^\dagger$}
{Email address:
giddings@physics.ucsb.edu}
}
\vskip.1in
\centerline{\sl Department of Physics}
\centerline{\sl University of California}
\centerline{\sl Santa Barbara, CA 93106-9530}

\bigskip
\centerline{\bf Abstract}

The large radius limit in the AdS/CFT correspondence is expected to
provide a holographic derivation of flat-space scattering amplitudes.
This suggests that questions of locality in the bulk should be
addressed in terms of properties of the S-matrix and their translation
into the conformal field theory.  There are, however, subtleties in
this translation related to generic growth of amplitudes near the
boundary of \ads\ space.  Flat space amplitudes are recovered 
after a delicate projection of CFT correlators onto normal-mode
frequencies of AdS.  Once such amplitudes are obtained from the CFT,
possible criteria for approximate bulk locality include bounds on
growth of amplitudes at high energies and reproduction of
semiclassical gravitational scattering at long distances.

\Date{}
%\draft

\newsec{Introduction}

Maldacena's proposed correspondence\refs{\Mald} between string (or M)
theory on $AdS_5\times S^5$ and $\caln =4$ supersymmetric gauge theory
in four dimensions has stimulated a great deal of recent
excitement.\foot{For a recent review and extensive set of references,
see \refs{\AGMOO}.}  A particularly fascinating aspect of this
correspondence is that it apparently serves as a concrete realization
of holographic ideas \refs{\tHoo,\Suss}.  Although a great deal of
work has been done to deduce properties of the large-$N$ Yang-Mills
conformal field theory from this correspondence, an even more
interesting question is how to deduce properties of string/M theory
from the the boundary conformal field theory.

Several steps have recently been taken in this direction.  In
particular, in \refs{\BGL,\bsm} the relationship between boundary
correlators and an AdS analog of the S-matrix (called the boundary
S-matrix in \bsm) was described.  This work had been closely preceded
by related work \refs{\Polc,\SussS} which sketched a prescription to
derive flat space S-matrix elements in the infinite radius $R$ limit
of AdS space; one naturally expects this S-matrix to be an appropriate
$R\rightarrow \infty$ limit of the boundary S-matrix of
\refs{\BGL,\bsm}.

This paper will investigate this question:  in particular, it will address
the issue of how flat-space S-matrix elements can be obtained from
conformal field theory data.  As we will find, this is somewhat nontrivial.

A useful analogy to bear in mind is that between \ads\ space and a
resonant cavity.  If one quantizes a free field in AdS, generic
frequencies produce non-normalizable states, and the normalizable
states correspond to a discrete set of frequencies and are analogous
to cavity modes.  Since there are no true asymptotic states among
these modes, it's not a priori clear how to formulate scattering
problems.  Here the example of a resonant cavity serves as a guide.
Consider for example two atoms at diametrically opposite ends of the
cavity, and suppose one is in an excited state, and one in the ground
state.  Even when the transition energy is not a normal-mode frequency
of the cavity, it is possible for atom one to decay to its ground
state with atom two transitioning to the corresponding excited state.
Or one may have more atoms, acting as emitters and detectors, and the
emitted particles may mutually scatter before being detected.  Such
Gedanken experiments give a framework to discuss
scattering.\foot{Another possibility in the resonant cavity case is to
cut small holes in the wall of the cavity through which particles may
enter and exit.  This clearly more difficult for AdS space, although
an analogous construction exists for the case where an AdS bubble is
embedded in a space with asymptotic particle states, and these states
must penetrate a potential barrier to reach the interior of AdS, as
described in \refs{\BGL,\bsm}.}

This analogy is not perfect.  For one thing, AdS space has much more
volume at infinity than flat space, and this together with growth at
infinity of the wavefunctions at non-normalizable frequencies can have
important consequences.  In particular, as we'll see, these combined
effects can lead to growth of interaction strengths near the AdS
boundary if we work with states at non-normalizable frequencies.  This
poses a difficulty for extracting flat-space scattering amplitudes in
the center of a large radius \ads\ space.  An obvious retort is that
one should project scattering amplitudes onto normalizable
frequencies, with a corresponding projection on the correlators of the
boundary CFT.  However, this is not made any easier by our lack of
knowledge of the spectrum of normalizable states at the multiloop or
non-perturbative levels, except in the case of protected states.  We
will discuss this problem and possible resolutions in more detail.

If we assume that it is indeed possible to extract flat-space
S-matrices from the conformal field theory, another profound question
arises.  If the correspondence of \Mald\ is correct, the theory in the
bulk should exhibit approximate $d+1$-dimensional locality, in an
appropriate low-energy limit.  Although the bulk theory is conjectured
to be string theory, which is manifestly non-local, this non-locality
is expected to only be apparent ``at the string scale'' (in some
appropriate sense), or perhaps in black hole experiments.  The theory
should be {\it macroscopically} local, namely it should give
low-energy, weak field amplitudes that are derivable from a local
low-energy effective field theory.  An important question is what
property of the boundary CFT implies macroscopic bulk locality in the
large radius limit of \ads\ space?  Also, one would like to better
characterize and understand the nonlocalities, and their possibly
quite important implications.

To address this question we need a way to diagnose locality.  From at least
two viewpoints (gauge invariance, holography) it was expected that the
AdS/CFT correspondence would only yield S-matrix elements -- not off-shell
amplitudes -- and direct study has partially confirmed this
\refs{\Polc,\SussS,\BGL,\bsm}.  Therefore we need a way of inferring from
the S-matrix whether -- or to what extent -- the underlying theory is
local.  Various bounds exist for S-matrices derived from local
theories, or even from generalizations to theories with non-local
behavior above a definite energy scale.  However, derivation of these
bounds becomes problematic in the case of theories with massless
particles, and particularly for theories with gravity.  There is some
related information about the high-energy structure of string
scattering amplitudes, but at present nothing that serves as a
definitive test of locality near the string scale.  At macroscopic
scales, one important test of locality is that the theory correctly
produce semiclassical gravity amplitudes or other coulombic behavior,
at long distances or equivalently small momentum transfer.

Yet another question that can be addressed is that of whether, or to what
extent, the underlying bulk theory can be reconstructed from complete
knowledge of the S-matrix elements.  This is a difficult problem, and in
general the solution is not unique.  A few comments about this problem will
also be made.

The paper will begin with a brief review of the Maldacena conjecture,
as formulated by Gubser, Klebanov, Polyakov, and Witten.  The next
section will then describe the basic properties of the large radius
limit for AdS.  Section four will then turn to the question of
whether, and how, bulk flat-space S-matrix elements can be extracted
from boundary conformal field theory correlators in the large radius
limit.  If the latter are written in frequency space, an important
distinction occurs between frequencies corresponding to normalizable
and non-normalizable modes; at frequencies corresponding to
non-normalizable modes, boundary correlators receive important
contributions from regions near the boundary of \ads\ space.  This
behavior poses some difficulty for extracting the flat-space S-matrix,
although outlines of a  procedure will be given.  Section
five contains some discussion of the problem of investigating the bulk
locality properties of the theory, as well as on the problem of
reconstructing the bulk theory.  Following the conclusion is a rather
lengthy Appendix containing a number of useful properties of \ads\
space, its propagators, and the large radius limit.  Many of these
appear previously in the literature, although there are some new
results.

\newsec{Review of the GKPW correspondence}

We begin by recalling the precise form of the correspondence for local
correlators, as formulated by Gubser, Polyakov, Klebanov and Witten
\refs{\GKP,\Witt}, and extended to lorentzian signature in
\refs{\BKL,\BKLT}.  This paper will use global coordinates
$x=(\tau,\rho,\Omega)$
for the cover of \ads\ space,
\eqn\glocoor{ds^2 = {R^2\over \cos^2\rho} ( - d\tau^2 +  d\rho^2 +
\sin^2\rho \, d\Omega^2_{d-1} )\ .
}
The boundary has topology $S^{d-1}\times {\bbb R}$, or alternatively can be
represented as the  conformally
equivalent infinite-sheeted cover of Minkowski space.  A point on the
boundary sphere can be specified by a $d$-dimensional unit vector $\ehat$;
boundary points are parameterized as 
$b=(\tau,\ehat)$.

The basic statement of this correspondence is 
\eqn\adscorr{Z[\phi(\phi_0)] = \langle T \, e^{i\int_\partial db \, \phi_0(b)
\, \calo(b)}\rangle\ .}
In this formula labels for different fields and their corresponding
operators have been suppressed.  On the left hand side is the bulk
partition function for string theory on $AdS_5\times S_5$ with both
radii set to $R$.  This is evaluated with bulk fields $\phi$ constrained to
obey the boundary condition
\eqn\fieldbc{\phi\limrho (\cos\rho)^{2h_-} \phi_0(b)\ }
where the constants $h_{\pm}$ and $\nu$, defined by
\eqn\hpmdef{4h_{\pm} = d \pm \sqrt{d^2 +4m^2R^2} = d\pm 2\nu\ ,}
for a field of mass $m$, govern the asymptotic behavior of the field.
The right side is the corresponding generating functional for CFT
correlation functions in ${\cal N}=4$ SU(N) Yang-Mills theory. The
operator $\calo$ corresponding to $\phi$ has conformal weight
$\Delta=2h_+$.  The parameters of the theories are related by 
$g_s=g_{YM}^2$ for the couplings, and $R=(g_{YM}^2 N)^{1/4}$ 
for the  $AdS_5\times S_5$ radius.

At the level of correlation functions, the
correspondence \adscorr\ may be rewritten as
\eqn\corrcorr{\langle T(\calo(b_1)\calo(b_2)\cdots\calo(b_n))\rangle =
\int \prod_{i=1}^n [dx_i \gbdel(b_i,x_i)] G_T(x_1,\cdots,x_n)\ .}
In this expression $\gbdel(b_i,x_i)$ denotes the full (multiloop)
bulk-boundary propagator, and $G_T(x_1,\cdots,x_n)$ denotes a bulk
$n$-point Green function with its external legs truncated.  
Some properties of bulk and bulk-boundary propagators will be reviewed in
the Appendix.

In particular, the basic issues will be illustrated using the 
relation between four-point functions,
\eqn\gkpwf{\langle T \calo_1(b_1) \cdots \calo_4(b_4)\rangle = \int
\prod_{i=1}^4 \left[dx_i \gbdel(b_i,x_i)\right] G_T(x_1,\ldots,x_4) \ .}
As a concrete example, consider truncating string theory down to a
scalar sector with a three-point coupling, as in \refs{\FMMR}, and
suppressing dependence on the $S^5$ coordinates:
\eqn\scaact{S= -\int dV  \left[\hf (\nabla \Phi)^2 + {m^2\over 2}
\Phi^2  + g \Phi^3\right]\ .}
%
%\eqn\scaact{S= -\int dV \left\{ \sum_i \left[\hf (\nabla \Phi_i)^2 + {m_i^2\over 2}
%\Phi_i^2 \right] + \sum_{ijk} g_{ijk} \Phi_i\Phi_j\Phi_k\right\}\ .}
%
To leading order in the coupling, the CFT correlator is given by a
term of the form
\eqn\Samp{\eqalign{\langle T \calo_1(b_1) \cdots \calo_4(b_4)\rangle_{{\rm
tree},t}  = &
- g^2 \int dV dV' \Big[\kbdel(b_1,x) \kbdel(b_3,x) K_B(x,x')\cr &
\kbdel(b_2,x') \kbdel(b_4,x')\Big]}}
plus $s$ and $u$ channel contributions, where $K_B$ is the bulk
Feynman propagator for the $\Phi$ field, and $\kbdel$ is the
tree-level bulk-boundary propagator.

\newsec{The large-$R$ limit}

Our goal is to recover flat-space S-matrix elements from boundary
correlators.  In order to extract these we must consider the regime of
large $R$, where 
the AdS geometry has a large region (of size $\calo$(R))
that is approximately flat.  

The relation between the large-$R$ geometry of AdS and flat space is
easily exhibited in the global coordinates \glocoor.  An arbitrary
point $P$ of AdS space can be moved to the origin $(\tau,\rho) =
(0,0)$ by an $SO(2,d)$ transformation.  For large $R$, we can readily
recover the nearly flat metric in the vicinity of $P$ in spherical
polar coordinates by defining
\eqn\minkcoord{t=R\tau\ ;\  r=R\rho\ .}  
Then
the metric \glocoor\ becomes
\eqn\appflat{ds^2 = {1\over \cos^2(r/R)}\left[ -dt^2 + dr^2 + R^2
\sin^2({r\over R})^2 d\Omega^2\right]\ .}
For $r\ll R$, this clearly reduces to the flat metric.

Since the spacetime in the vicinity of $P$ is approximately flat, one
expects to recover flat-space physics in this region.  For example,
consider the bulk Feynman propagator, which appeared in the amplitude
\Samp.  This has been given in closed form in
terms of a hypergeometric function\refs{\BuLu, \InOo}, and the large $R$ limit
of this expression is derived in the Appendix.  The result is, for
particles\foot{The case of massive particles, $m\gg  \calo(1/R)$,
can also be treated.} of masses  $\roughly< \calo(1/R)$, 
\eqn\limprop{i K_B(x,x') = {{\tilde C}\over [s^2(x,x')+ 
i\epsilon]^{(d-1)/2}}\ ,}
where $\tilde C$ is a constant given in the Appendix, and $s(x,x')$ is
the geodesic distance between the points $x$ and $x'$.  This is the
standard flat-space propagator for a massless field.

\newsec{S-matrices at large $R$?}

We'd like to determine whether the flat-space S-matrix can be extracted
from the conformal correlators, such as \Samp, in the limit of large $R$.  
To begin with, recall the form of the flat-space S-matrix for the process
corresponding to that of \Samp.  This is simply the 
tree-level, $t$-channel contribution to the flat-space
S-matrix, which is given by a Fourier transform of the Feynman propagator:
\eqn\Stf{\eqalign{S_{{\rm Mink},t}(s,t) &=-g^2\int d^{d+1}x d^{d+1}x'
e^{i(k_1+k_3)\cdot x} e^{i(k_2+k_4)\cdot x'} K_F(x,x')\cr & =-ig^2 
\delta^{(d+1)}(k_1+k_2+k_3+k_4) {1\over t}\ .}}

The general question at hand is how to extract the 
S-matrix, for example \Stf,   
directly from the corresponding contribution to the CFT correlation
function, for example \Samp.  Note that these formulas appear very similar.  By the 
LSZ prescription, the flat S-matrix consists of truncated flat-space $n$-point
functions, convoluted against on-shell wavefunctions; this should be
compared to \Samp, which has the identical structure with bulk-boundary
propagators replacing flat-space wavefunctions.  This served as a
central observation behind the
interpretation of the conformal correlators as providing a ``boundary
S-Matrix'' for \ads\ space\refs{\BGL,\bsm}.  

In searching for a general
prescription to extract the flat-space S-matrix from the AdS/CFT
correspondence, we will begin by ``reverse engineering'' the expression
\Samp\ to see how the corresponding piece of the 
S-matrix could be extracted from this contribution to the conformal
correlators.  This will allow us to infer some lessons for the more
general problem of the complete S-matrix.

We've just seen that at large $R$ the contribution from $K_B$ to the
correlator reduces to the expected flat-space expression.  Therefore,
the remaining task is to understand whether the contributions from the
factors $\kbdel$ can be related to on-shell, flat-space wavefunctions
in the Minkowski region.  Some properties of $\kbdel$ are reviewed in
the appendix.  Given the relationship \minkcoord, it is convenient to
work with the frequency conjugate to the global AdS time $\tau$, as
this also corresponds to definite Minkowski energy,
\eqn\enerrel{\omega=ER\ .}

In frequency space, the relation \Samp\ becomes
\eqn\fourpt{\eqalign{\langle
T\calo(\omega_1,\ehat_1)\cdots\calo(\omega_4,\ehat_4)\rangle_{{\rm tree},t}
 =& -g^2 \int
dV  dV'\Big[ \kbdel(\omega_1,\ehat_1;x)
\kbdel(\omega_3,\ehat_3;x)\cr &  K_B(x,x') 
\kbdel(\omega_2,\ehat_2;x')
\kbdel(\omega_4,\ehat_4;x')\Big] \ .}}
It will turn out that there is an important distinction between the cases
where the frequency of the external state is generic, corresponding to a
non-normalizable mode, and where it is that of one of the normalizable
modes.  We will consider these in turn.

\subsec{Generic frequencies}

In the case where the $\omega_i$ don't correspond to normal-mode
frequencies, 
the
bulk-boundary propagator must be a non-normalizable solution of the scalar
wave equation.  One can most easily explore the consequences of this by
working in a partial-wave basis; as seen from \appflat\ (and also shown
in terms of generators in the Appendix), AdS angular momentum is
directly identified with Minkowski angular momentum in the large-$R$ limit.
In an angular momentum basis and at arbitrary frequency, the bulk-boundary
propagator has asymptotic behavior 
\eqn\bbasy{\kbdel(\omega,l,\vecm;x)\rightarrow e^{i\omega\tau} Y^*_{l\vecm}
(\cos\rho)^{2h_-}\ .}
From this we immediately see a problem in extracting the flat-space
S-matrix, \Stf, directly from \fourpt.  In order to do so, we'd like $K_B$ to
be convolved with wavefunctions that have their support concentrated in
the flat region $r\ll R$.  However, the behavior \bbasy\ ensures that all
linear combinations of the non-normalizable modes grow like $(\cos
\rho)^{2h_-}$ at infinity, and thus are concentrated in the region $r\gg R$
instead.  Indeed, consider the behavior of the integral over $x$ in
\fourpt.  
The
volume element is
\eqn\volelem{dV = {R^{d+1} (\sin\rho)^{d-1}\over (\cos\rho)^{d+1}} 
d\tau d\rho d^{d-1} \Omega\
,}
%
%and the bulk Green function behaves asymptotically as
%
%\eqn\gasy{K_B(x,x')\rightarrow (\cos\rho)^2h_+ G(b,x')\ .}
%
and so near $\rho=\pi/2$ the $\rho$ part of the integral takes the form
\eqn\asympint{\sim \int d\rho { (\sin\rho)^{d-1}\over (\cos\rho)^{d+1} } 
(\cos\rho)^{4h_-}
K_B(x,x')\sim \int d\rho {1\over (\cos\rho)^{2\nu+1}} K_B(x,x')\ .}
Thus the wave-function factor convolving the bulk Green function has
its main support in the vicinity $\rho\approx \pi/2$.\foot{This is
directly connected to the fact that, in the more general case of
scalar fields of unequal masses, the integral \asympint\ only
converges for certain values of the scalar masses\refs{\FMMR}.}  For
non-normalizable frequencies, there is no apparent way to obtain a
limit in which one recovers the desired asymptotic falloff of
flat-space wavefunctions for $r$ large but $r\ll R$ without also
encountering this growth at infinity.

Evidence for this behavior can also be seen directly in position
space.  As shown in the Appendix, the position space bulk-boundary
propagator takes the form 
\eqn\gbbpos{\kbdel(b,x')= C_{B\partial} \left[{\cos\rho'\over
\cos(\tau-\tau') - \sin\rho' \ehat\cdot\ehat' }
\right]^{2h_+}\ }
up to the $i\epsilon$ prescription (see Appendix).
This is singular for points on the boundary light cone: at points such
that $cos(\tau-\tau') - \ehat\cdot\ehat'=0$, it has behavior $\sim
(\rho-\pi/2)^{-2h_+}$ at $\rho\rightarrow\pi/2$.  In \Samp, the
product of the first two bulk-boundary propagators will be singular on
the intersection of their boundary light cones; this behavior is
exacerbated by the growth of the volume element.

These arguments, suggest that for generic frequencies AdS
holography is in a sense only ``skin deep.''

\subsec{Normalizable frequencies}

We now turn to the case where all the external frequencies $\omega_i$
correspond to those of normalizable modes.  To investigate this case,
recall that the bulk-boundary propagator can be found as a limiting case of
the bulk propagator,\foot{For more details, see the Appendix.}
\eqn\gbbdef{\gbdel(b,x') = 2\nu i R^{d-1}
\lim_{\rho\rightarrow\pi/2} (\cos\rho)^{-2h_+}
G_B(x,x')\ ,}
and that the Feynman propagator can be written in the form
\eqn\feyndef{iK_B(x,x')=\int {d\omega\over 2\pi} \sum_{nl\vecm}
e^{i\omega(\tau-\tau')} {\phi^*_{nl\vecm}(\rho,\ehat)
\phi_{nl\vecm}(\rho',\ehat')\over \omega_{nl}^2-\omega^2 -i\epsilon}\ .}
The normalizable wavefunctions $\phi_{nl\vecm}(\rho,\ehat)$ have asymptotic
behavior
\eqn\normlim{\phi_{nl\vecm}(\rho,\ehat) \limrho k_{nl} (\cos\rho)^{2h_+}
Y_{l\vecm}(\ehat)}
for certain constants $k_{nl}$ (see Appendix).  
The frequency-space form of $\kbdel$ is thus
\eqn\gbbsum{\kbdel(\omega,\ehat;x') = 2\nu R^{d-1}
e^{i\omega\tau'} \sum_{n,l,\vecm}{k_{nl} Y^*_{l\vecm}(\ehat)
\phi_{nl\vecm}(\rho',\ehat')\over \omega_{nl}^2-\omega^2 -i\epsilon}\ .}

From \gbbsum\ we see that extracting the normalizable-frequency piece
is rather delicate.  We want only the contribution corresponding
precisely to the normalizable-mode frequency, in order to eliminate
the above non-normalizable behavior.  The Green function is divergent
at this frequency; one must extract the residue at the pole.  Once one
takes into account higher-loop corrections, these frequencies are not
a-priori known (though they should be determined by knowledge of the
exact conformal weights), and the frequency-space behavior of the
amplitude may be more complicated.  These factors pose significant
difficulties.  Nevertheless, let us assume that we are able to perform
these steps, and see where they lead.

Thus, we
define a modified bulk-boundary propagator by 
\eqn\newbb{\eqalign{\khat(n,l,\vecm,x)&=  \lim_{\omega\rightarrow\omega_{nl}}
(\omega^2-\omega_{nl}^2) \kbdel(\omega,l,\vecm, x)\cr 
&= 2\onl\oint{d\omega\over 2\pi i} \kbdel (\omega,l,\vecm,x)\cr
&=-2\nu R^{d-1} e^{i\omega_{nl}\tau} k_{nl}
\phi_{nl\vecm}^*(x)\ .}}
The corresponding operations can be performed directly on CFT
correlators, where they are the operations needed to project onto a
definite {\it state} of the CFT.

The modified propagator \newbb\ does reproduce on-shell wavefunctions in
flat space.  Indeed, in the appendix it is shown that in the $r\ll R$
limit, the modes $\phi_{nl\vecm}$ go over into the flat-space
wavefunctions,
\eqn\limmodes{\phi_{nl\vecm}(\vecx) \limR \sqrt{2E} {1\over r^{d/2-1}}
J_{l+d/2-1}(Er) Y_{l\vecm}(\ehat)\ .}
Likewise, the coefficients $k_{nl}$ can be worked out, with the result
\eqn\newlim{\khat(n,l,\vecm,x)\limR C(E,R) 
i^{l} {1\over (Er)^{d/2-1}} J_{l+d/2-1}(Er) Y_{l\vecm}^*(\ehat)\ }
with coefficient function
\eqn\Cer{ C(E,R)=
-{2^{2-\nu} \over \Gamma(\nu) } (-1)^{ER/2-h_+}
(ER)^{2h_+}\ .}
In fact, this can easily be transformed back to position space on the
boundary, giving\foot{Note that the following is valid to the extent
that contributions from very large $l$ don't contribute to the sum
over all $l$.  Such contributions give AdS corrections to the plane
waves.  These can be suppressed by consideration of wavepackets with
spatial spread $<R$.}
\eqn\newbbp{\khat(E,\ehat;x) = \sum_{l,\vecm} Y_{l\vecm}(\ehat)
\khat(n,l,\vecm,x) \limR { C(E,R)\over (2\pi)^{d/2}}  e^{i\veck\cdot\vecx}\ ,}
a plane wave with $\veck = E\ehat$. 

Putting all of this together suggests tree-level prescriptions for extracting
flat-space S-matrix elements from boundary correlators:
\eqn\flsmat{\eqalign{ S[&k_1,\cdots,k_n]\cr &=\prod_{i=1}^n
\left[  (2\pi)^{d/2} 2E_i R C(E_i,R)^{-1} \oint_{E_iR} 
{d\omega\over 2\pi i} \right] \langle T\calo(\omega_1,\ehat_1)\cdots\calo(
\omega_n,\ehat_n)\rangle\cr
&= \prod_{i=1}^n
\left[ (2\pi)^{d/2}  C(E_i,R)^{-1}
\lim_{\omega_i\rightarrow \omega_{n_il_i}} (\omega_i^2 -
\omega_{n_il_i}^2)\right]
\langle T\calo(\omega_1,\ehat_1)\cdots\calo(\omega_n,\ehat_n)\rangle
 }}
where $\omega_{n_il_i} = E_i R$ is a normal mode frequency, and $k_i=E_i(1,\ehat_i)$.

However, at the multiloop level these expressions are potentially
problematical.  In particular, one doesn't a-priori know the
frequencies to tune $\omega_i$ to in order to sit on a normalizable
mode, and the analytic structure in $\omega$ may contain more than
just simple poles at these frequencies.  One possible approach to this
problem -- given complete knowledge of the Yang-Mills correlators --
would be to take a correlator and look for the frequencies where poles
appear, extract the residues at these poles, and then use the result
to construct the flat-space S-matrix along the above lines.  These
frequencies  are approximately given by locating the poles in the
boundary two-point function, which at tree level takes the form (see
Appendix)
\eqn\bdytwo{\eqalign{K_\partial(b,b')&
 \propto\lim_{\rho,\rho'\rightarrow \pi/2}
(\cos\rho\cos\rho')^{-2h_+} K_B(x,x')\cr & = k_\partial
\int {d\omega\over 2\pi} e^{i\omega(\tau-\tau')} 
\sum_{nl\vecm} {k_{nl}^2 Y^*_{l\vecm}(\ehat) Y_{l\vecm}(\ehat')\over 
\omega_{nl}^2-\omega^2}
%\delta_{ll'}\delta_{\vecm\vecm'} \delta(\omega-\omega')
\ .}}
However, interactions will in general shift the energy of the
two-particle state relative to twice the single-particle energy,
causing added difficulty in precisely identifying the relevant frequencies.

Another approach, advocated in \refs{\Polc,\SussS}, is to convolve the
boundary correlators with appropriately chosen wavepackets.  This
appears to have some difficulty, as we can see from \gbbsum.  If
$f(\omega,\ehat)$ is the wavepacket profile for one of the external
states, then it is connected to the rest of the diagram through a
factor of the form
\eqn\wavetr{\eqalign{\int d\omega d^{d-1}\Omega& f(\omega,\ehat)
\kbdel(\omega,\ehat,x') \cr &= \int d\omega d^{d-1}\Omega f(\omega,\ehat) 
\sum_{n,l,\vecm}2\nu R^{d-1}{k_{nl} Y^*_{l\vecm}(\ehat)
\phi_{nl\vecm}(\rho',\ehat')\over \omega_{nl}^2-\omega^2 -i\epsilon}
e^{i\omega\tau'}\ .}}
For any regular $f$ this receives contributions from non-normalizable
frequencies even if $f$ is sharply peaked near a specific normalizable
frequency.  This in turn implies sensitivity to contributions from
interactions at $r\gg R$, as described in Sec.~4.1.  It is not clear
how to make wavepackets focussed in the Minkowski region from these
modes.

A modification of this procedure would be to {\it first} extract the
amplitudes restricted to normalizable frequencies, as outlined above.
Then wavepackets can be built by taking linear combinations of those
with different normal mode frequencies.

These results suggest that while it may be possible to derive
flat-space S-matrix elements from the AdS/CFT correspondence, the
procedure is not so simple as it first appeared.

\newsec{Locality}

Let us assume that it is possible to infer the full flat-space
S-matrix from the conformal field theory correlators , through some
variant of the above procedure or some other procedure.  This would
provide an even more concrete realization of the holographic proposal.
A profound question underlying this proposal is how it is possible for
the boundary theory to produce a bulk theory that is an approximately
local theory in one higher dimension.

One piece of the answer appears to be that the boundary theory is a
conformal field theory.  In a conformal field theory, there is no
mass-shell condition on the states, since there are no masses.
Correspondingly, one can have states with fixed boundary momenta and a
spectrum (discrete in global coordinates, continuous in Poincare
coordinates) of frequencies.  This statement is made manifest in the
Kallen-Lehman representation for the boundary two-point function,
which can be found from that of AdS given in \refs{\DuFr}.
In the bulk theory this spectrum is interpreted as arising from the
different values of the momentum in the extra radial direction.

Thus, in this sense the boundary theory  contains enough states to
represent a bulk theory.  However, this is no guarantee that the
interactions have the correct properties to produce a sensible and
approximately local bulk theory.  One other key property is bulk
momentum conservation, which has been the subject of one recent
discussion\refs{\Li}.  This seems assured by the correspondence
between the symmetries of the two theories: the conformal group
$SO(2,d)$ is also the group of isometries of \ads\ space, and in the
large-$R$ limit reduces to the Poincare group.  However, this does not
imply locality -- there is an infinite variety of momentum conserving
but nonlocal interactions.

One would like to investigate the possible presence of such
nonlocality in the bulk theory.  Of course, the bulk theory is not
expected to be a local theory, but rather to have nonlocalities
present on a scale of order the string scale.  But this is to be
contrasted with the situation where there are macroscopic
nonlocalities, for example on scales of order the (large) AdS radius.
One could imagine an observer living in an approximately Minkowski
region in a very large AdS space without ever knowing about the large-scale
curvature, and to such observers the only nonlocalities present
should be very subtle and difficult to measure effects not easily seen
at long distances.  Such an AdS observer shouldn't be able to exploit
macroscopic nonlocality to  win the lottery!  What
property is it of the boundary conformal theory that ensures
preservation of locality at the macroscopic level?  How does one
characterize the amount of nonlocality present?  What experiments
could be performed to measure it?  And is it sufficient to resolve the
black hole information paradox or solve the cosmological constant
problem?  These are all questions of considerable importance.

As expected on grounds of both holography and gauge invariance, it
appears that one can at best compute the S-matrix of the bulk theory
from full knowledge of the CFT correlators.  Therefore conventional
tests of locality -- like commutativity of field operators at
spacelike separations -- aren't available.  One must find ways of
diagnosing locality directly from the S-matrix.

This is a difficult problem.\foot{I thank K. Bardakci for several
conversations on this issue.}  First consider theories with a mass
gap.  Here one test of locality for the S-matrix is that it respect 
various bounds that can be derived as a result of locality.
Amplitudes must satisfy both upper and lower bounds.  For example, the
Froissart bound
\refs{\Froi} states that for four-point scattering at arbitrary angle,
the amplitude must obey
\eqn\frois{|A(s,\theta)|< C s \log^2 s\ ,}
where $C$ is a constant; there are more stringent bounds for fixed
angle, $0<\theta<\pi$.  Polynomial boundedness also implies the
Cerulus-Martin bound \refs{\CeMa,\Mart}, which states that amplitudes at
arbitrary angles can't fall too rapidly at fixed angle:
\eqn\ceru{ |A(s,\theta)| \geq s^{-c\sqrt s}\ .}

Similar bounds have also been found in nonlocal theories.  One way
of producing a nonlocal theory is to construct a theory with an
exponentially increasing density of states, such as string theory.
This leads to the definition of quasi-localizable theories,\foot{For
more discussion of these and their bounds, see \refs{\FaSo} and references
therein\ .} which are theories with nonlocality occurring below a
definite length scale $l$.  These theories are characterized for example by
densities of states that grow like
\eqn\dengrow{\rho(m)\sim e^{ (lm)^\gamma}}
for some power $\gamma$,
and satisfy bounds of the form
\eqn\qlbd{ |A(s,\theta)| < C' s^2 \log \rho( s^{1/2})\ .}
This suggests the possibility of reading off the scale of nonlocality
from the behavior of the four-point scattering amplitude.

However, derivation of these bounds assumes the existence of a gap.
The requisite analyticity properties are spoiled by massless
particles; gravity is particularly problematical.  Here of course the
usual IR divergences imply that in four dimensions one must study inclusive
cross-sections, summing over soft particles below some energy
resolution relevant to the experiment in question.  The S-matrix  can
be defined in higher dimensions.\foot{I thank T. Banks for a conversation
on this point.}  One might hope that similar bounds could be derived for
either the 4d inclusive rates or for higher-dimensional S-matrices,
providing a way of quantifying the degree of nonlocality of
the theory.\foot{Also, in AdS the radius supplies an intrinsic IR
cutoff, which may be useful in circumventing the usual IR problems.}
This is an important problem for the future.

We have some partial information about high-energy string scattering\refs{\GrMe,
\Amat} that might  be considered as a standard of comparison.
For example, \GrMe\ investigated the large $s$, fixed angle
regime, and found perturbative amplitudes of the form
\eqn\borel{|A_G(s,\theta)|\sim e^{-(s\ln s + t\ln t + u\ln u)/4G}}
at genus $G$.  
It would be very interesting to see whether the CFT reproduces this
and other stringy behavior.  Moreover, 
there are a number of open
questions about the large order/nonperturbative completion of results such
as \borel\ for 
high-energy string scattering, and one hope is that the boundary CFT
could teach us something new about this.

So far the discussion has focussed on locality in the large $s$
regime.  However, it's not even a-priori clear how the AdS/CFT
correspondence produces bulk locality at the macroscopic level.  
We'd like to find appropriate criteria for this.

One possibility is to for example consider fixed energy scattering of
physical particles\foot{These must in particular be neutral under any
nonabelian gauge groups.} at large distances, or equivalently small
$t$.  One very rough criterion is that we have interactions falling at
least as fast as $1/r^{d-2}$ at long distances, since the only long
range forces are expected to be gravity and other coulombic
interactions.  This corresponds to scattering amplitudes that grow
like $ 1/t$ as t decreases.  In the AdS context, one of course expects
this growth to be truncated at the AdS radius, $t\sim (1/R)^2$, but
for $R\gg1$ there is a clean separation of scales and one can study
growth of amplitudes in the regime $1\gg t \gg (1/R)^2$ to see if they
satisfy this crude criterion.  Indeed, taking this one step further,
one could also ask whether the boundary theory reproduces the correct
structure for gravitational (or other coulombic) scattering for a wide
variety of semiclassical states.  This would be an important test,
sensitive to bulk locality and more, of any independent calculation of
the CFT correlators.  It would be very interesting to go beyond to
understand what property of the boundary theory ensures recovery of
the correct semiclassical limit.

One might inquire whether locality properties are
encoded nicely in the operator product expansion of the boundary
theory.  To investigate this, one first needs a translation of the
kinematical variables -- the Mandelstam invariants -- into the AdS and
CFT contexts.  This is provided by considering the quadratic Casimir
of SO(2,d), which in the large-$R$ limit reduces to the Poincare
invariant $P^2$ as shown in the appendix.  Thus if we wish to combine
two states in representations of SO(2,d), the analog of $s$ is now
provided by the conformal weight $\Delta$ of the resulting states in
the product representation.

First consider the problem of large-$s$ scattering.  This for example
should be governed by fusion of two high energy states into a third
state with large $\Delta.$  The OPE takes the general form
\eqn\ope{ \phi_{\Delta_i}(x) \phi_{\Delta_j}(y) \sim 
\sum_k c_{ijk} {\phi_{\Delta_k}(y) \over |x-y|^{\Delta_i + 
\Delta_j - \Delta_k}}\ .}
Thus large $\Delta_k$ corresponds to very high order terms in the OPE,
rather than the leading behavior.  One can see a similar effect by
considering scattering of two particles with momenta $E(1,\ehat_1)$
and $E(1,\ehat_2)$.  For these, $s = E^2(1-\ehat_1\cdot\ehat_2)$.  In
the OPE limit, $\ehat_1\rightarrow \ehat_2$, $s\sim E^2 \theta^2$
where $\theta$ is the angle between the unit vectors.  Small boundary
distance only corresponds to large $s$ if we simultaneously take very
large $E$.

However, this suggests that behavior of amplitudes at $t\rightarrow0$
could be explored in the OPE limit in the $t$ channel.  I hope to
return to the implications for CFT in future work.

Finally, related to this discussion is the question of whether one can
reconstruct the entire bulk theory given complete data in the boundary
theory.  In general reconstruction of a theory given the S-matrix is
not unique, but one may try to reconstruct even {\it one} bulk theory
that reproduces the correct amplitudes.  Consider first three-point
functions.  A problem here is that the CFT three-point function is
uniquely determined by conformal symmetry, up to a constant; in the
example of scalars, both interactions
\eqn\twoint{ g_1\int dV \phi^3\quad {\rm and}\quad 
g_2 \int dV \phi^2\partial^2\phi\ }
give identical three-point functions up to this constant
\refs{\FMMR}.  Of course the dependence of this constant on $R$ for
the two different cases will be different; this will help in decoding
the different interactions.  But in general the program will involve 
considering four- and higher-point functions, with their
non-trivial dynamics.  The problem of reconstruction is an interesting
one for the future.

\newsec{Conclusion}

This paper has attempted a modest beginning of an investigation of
bulk locality in the AdS-CFT correspondence.  A first problem is to
extract the flat space S-matrix from the boundary correlators.  Ideas
for how to do this have been previously presented in
\refs{\Polc,\SussS,\BGL,\bsm}, but there are subtleties.  In
particular, it was showed that at arbitrary {\it non-normalizable}
frequencies, the boundary correlators are sensitive to interactions in
the ``skin'' of AdS, at radii $r>R$, due to growth of the
wavefunctions and of the volume of \ads\ space.  It appears that only
by a delicate procedure of extracting the residues of poles at the
normalizable frequencies may we find the flat S-matrix.  There are
bound to be wrinkles in this procedure when all-order perturbative or
non-perturbative amplitudes are considered.

We'd like to know what form locality takes in the bulk theory.  There
is no known procedure to extract off-shell data about this theory --
in accord with holography as well as gauge invariance -- and so
information about locality properties must be derived directly from
the S-matrix.  In theories with a gap, locality or even nonlocality on
a definite scale implies certain bounds for the S-matrix, but the
presence of massless particles and particularly gravity complicates
the story.  Nonetheless, some information is known about the behavior
of high-energy string scattering amplitudes, and one might as a first
test try to investigate this behavior from the CFT and even to go
beyond to new results.  Furthermore, the bulk theory should exhibit
macroscopic locality, namely bulk observers should find an
approximately local theory at long distances.  One very rough
criterion for this is falloff of potentials (thus growth of
amplitudes) bounded by Coulomb behavior.  Indeed, another important
test is reproduction of semiclassical gravitational scattering.
We'd like to know whether the CFT
reproduces such behavior, and in particular what properties of the
boundary theory allow the surprising result that an approximately
local higher-dimensional theory emerges from it.

\bigskip\bigskip\centerline{{\bf Acknowledgments}}\nobreak

I have greatly benefited from conversations with K. Bardakci,
D. Gross, G. Horowitz, M. Kugler, D. Jackson, H. Ooguri, S. Shenker, and
J. Polchinski who has been working on closely related ideas.  I'd like
to particularly thank the theory group, and especially my host
S. Kachru, at Berkeley and Lawrence Berkeley Labs, where much of this
work was carried out, for their kind hospitality, and the Institute
for Theoretical Physics and Center for Mathematical Physics at the
University of Amsterdam, where this work was completed.  This work was
partially supported by DOE contract DE-FG-03-91ER40618.

\appendix{A}{Basic AdS tools}

This appendix will review some basic properties of the geometry of
\ads\ space and its boundary, and of wavefunctions and propagators on
AdS, as well as deriving some new and useful results.  In particular,
an explicit treatment of the large $R$ limit will be given.

\subsec{Geometry}

$AdS_{d+1}$ can be represented as the solution of the equation 
\eqn\adsdef{ (X^M)^2 = -R^2}
in flat $(2,d)$ signature Minkowski space with coordinates
$(X^{-1},X^0,X^i)$ and with metric
\eqn\minkmet{dS^2 = \eta_{MN}dX^MdX^N= - (dX^{-1})^2 - (dX^0)^2 + (dX^i)^2\ .}
Global coordinates $(\tau,\rho,\ehat)$, where $\ehat$ is a $d$-dimensional
unit vector, are defined by
\eqn\glodef{ X^{-1}= R{\cos\tau\over \cos\rho}\ ,\ X^0 = R{\sin \tau\over
\cos\rho}\ ,\ X^i = R\tan\rho \ehat^i\ ,}
and in these coordinates the metric takes the form
\eqn\glocoora{ds^2 = {R^2\over \cos^2\rho} ( -d\tau^2 + d\rho^2 +
\sin^2\rho \, d\Omega^2_{d-1} )\ .}
Typically we will work on the universal cover of \ads\ space, which in an
abuse of notation will also be denoted $AdS_{d+1}$.
The ${\bbb R}\times S^{d-1}$ boundary of this cover corresponds to
$\rho=\pi/2$ and is parameterized as $b=(\tau,\ehat)$.

There are two related notions of invariant distance on \mads.  The first is
{\it geodetic} distance, defined with respect to the embedding space metric:
\eqn\geodet{\sigma(X_1,X_2) = \hf\eta_{MN} \Delta X^M \Delta X^N\ ;}
in global coordinates it can be shown that
\eqn\geodeg{\sigma(x_1,x_2) = -R^2 + {R^2\over \cos\rho_1 \cos\rho_2}\left
[\cos(\tau_1-\tau_2) - \sin\rho_1 \sin\rho_2 \ehat_1\cdot\ehat_2\right]\ .}
The second is {\it geodesic} distance, as measured in the AdS metric
\glocoora.  This can be shown to be given by
\eqn\geodes{ s(x_1,x_2) = R \cosh^{-1} \left[{\cos(\tau_1-\tau_2)- 
\sin\rho_1 \sin\rho_2 \ehat_1\cdot\ehat_2 \over
\cos\rho_1\cos\rho_2}\right]}
in global coordinates.  Geodesic and geodetic distances are related by
\eqn\geogeo{\cosh\left({s\over R}\right) = 1 + {\sigma\over R^2}\ .}

There is of course no conformally-invariant notion of interval on the
boundary.  When working in Poincare coordinates for \mads
\eqn\poinmet{ds^2 = R^2\left( {dU^2\over U^2} + U^2 dx^2\right)\ ,}
where
\eqn\poincoor{U={X^{-1}-X^d\over R}\ ;\ x^\alpha = X^\alpha/RU\ ,\ 
\alpha=0,1,\cdots
d-1\ ,}
one frequently uses the non-conformally invariant interval $x_{12}^2=|x_1-x_2|^2$.
In global coordinates this becomes
\eqn\nonconf{x_{12}^2 = {2 (\cos(\tau_1-\tau_2) -\ehat_1\cdot\ehat_2)\over
(\cos \tau_1 - \ehat_1^d) (\cos \tau_2 - \ehat_2^d)} \ ;}
thus the definition 
\eqn\binter{s_{12}^2 =\cos(\tau_1-\tau_2) -\ehat_1\cdot\ehat_2}
gives an
analogous non-conformally invariant interval in the global frame.  In order
to form conformal invariants, one must have four or more points, allowing
the definition of cross ratios like
\eqn\crossdef{ {s_{12}^2s_{34}^2\over s_{13}^2s_{24}^2} =
{ (\cos(\tau_1-\tau_2) -\ehat_1\cdot\ehat_2) (\cos(\tau_3-\tau_4)
-\ehat_3\cdot\ehat_4) \over 
(\cos(\tau_1-\tau_3) -\ehat_1\cdot\ehat_3) (\cos(\tau_2-\tau_4)
-\ehat_2\cdot\ehat_4)}\ .}

\subsec{Wavefunctions}

In many respects \ads\ space behaves like a resonant cavity.  In particular,
solutions of the scalar wave equation
\eqn\scawav{(\sq - m^2)\phi=0}
are for generic frequencies non-normalizable, having asymptotic behavior
\eqn\asnon{\phi\propto (\cos\rho)^{2h_-}\ ,}
where
\eqn\hpmdefa{ 4h_{\pm} = d \pm\sqrt{d^2+ 4m^2R^2} = d\pm 2\nu\ ,}
near the boundary at $\rho=\pi/2$.  

Only for special frequencies,
\eqn\normfreq{\omega_{nl} = 2h_+ + 2n+l}
do normalizable solutions exist.  Explicit forms for these solutions
are given in \refs{\BuLu,\BKL}, 
\eqn\normmodes{\phi_{nl\vecm}(\vecx,\tau) = \chi_{nl}(\rho) Y_{l\vecm}(\ehat)
{e^{-i\onl\tau}\over \sqrt{2\onl}}}
where the radial wavefunctions $\chi_{nl}(\rho)$ are written in terms of 
the Jacobi
polynomials,\foot{The conventions of \refs{\AbSt} will be followed.}
\eqn\normmodes{ \chi_{nl}(\rho)
= A_{nl} (\cos\rho)^{2h_+}
(\sin\rho)^l P_n^{l+d/2-1,\nu}(\cos 2\rho) \ .}
Here $\Anl$ is a normalization constant.  If the $\phi_{nl\vecm}$ are given
the conventional Klein-Gordon normalization
\eqn\kgnorm{(\phi_{nl\vecm},\phi_{n'l'\vecm'})= \int d\Sigma^\mu
\phi_{nl\vecm}^* i{\overleftrightarrow \partial}_\mu
\phi_{n'l'\vecm'} = \delta_{nn'} \delta_{ll'} \delta_{\vecm\vecm'}\ }
with respect to surfaces of constant $\tau$,
then these constants are 
\eqn\anldef{\Anl^2= {2\onl \over R^{d-1}} {n! \Gamma(n+2h_++l)\over
\Gamma(n+l+{d\over 2}) \Gamma(n+\nu +1)}\ .}

The normalizable solutions have asymptotic behavior given by
\eqn\normasyma{\chi_{nl}(\rho) \limrho k_{nl} (\cos \rho)^{2h_+}\ ,}
with constants $k_{nl}$ given by
\eqn\knldef{k_{nl}= (-1)^n \Anl {\Gamma(n+\nu+1) \over n! \Gamma(\nu+1)}\
.}

\subsec{Green functions}

The bulk Feynman Green function is defined by solving
\eqn\feyneqn{(\sq-m^2)i K_B(x,x') = -\delta(x,x')}
with Feynman boundary conditions.  It can be represented as an infinite sum
over the normalizable modes,
\eqn\feynsum{iK_B(x,x') = \int {d\omega\over 2\pi} \sum_{\nlm}
e^{i\omega(\tau-\tau')} {\phi_\nlm^*(\vecx) \phi_\nlm(\vecx') \over
\onl^2-\omega^2 -i\epsilon}\ .}
The sum has been explicitly performed to yield\refs{\BuLu,\InOo}
\eqn\feyngreen{iK_B(x,x')=  {C_B\over 
[\cosh^2(s/R)]^{h_+}} F\left(h_+,h_++\hf;
\nu+1; {1\over \cosh^2(s/R)}-i\epsilon\right) \ ,}
where $C_B$ is a constant,
\eqn\Cdef{C_B={1\over R^{d-1}} {\Gamma(2h_+) \over 2^{2h_++1} \pi^{d/2}
\Gamma(\nu+1)}\ ,}
and $F$ is the hypergeometric function.
%
%\eqn\epsdef{\epsilon_{\tau-\tau'}=\cases{ \epsilon>0&if $\tau-\tau'>0$,\cr
%-\epsilon&if $\tau-\tau'<0$.\cr}\ .}
%

The bulk-boundary propagator $\kbdel$ is designed to provide a
solution to the free wave equation \scawav\ satisfying the boundary
condition
\eqn\phibc{\phi\limrho (\cos\rho)^{2h_-} f(b)\ .}
This solution is given by
\eqn\bbdefone{\phi(x) = \int db f(b) \kbdel(b,x)\ .}
A simple Green's theorem-type argument\refs{\bsm} shows that
\eqn\gbbdef{\kbdel(b,x') = 2\nu R^{d-1}
\lim_{\rho\rightarrow\pi/2} (\cos\rho)^{-2h_+}
iK_B(x,x')\ .}
We can therefore find two equivalent expressions for $\kbdel$, the first
from the limit of \feynsum,
\eqn\gbbsum{\kbdel(b,x') = 2\nu R^{d-1} \int {d\omega\over 2\pi} \sum_\nlm 
e^{i\omega(\tau-\tau')} {k_{nl}Y^*_{l\vecm}(\ehat) \phi_\nlm(\vecx') \over
\onl^2-\omega^2 -i\epsilon}\ ,}
and the second from the limit of \feyngreen,
\eqn\kbdele{\kbdel(b,x')= C_{B\partial} \left[{\cos^2\rho'\over
[\cos(\tau-\tau') - \sin\rho' \ehat\cdot\ehat']^2+i\epsilon }
\right]^{h_+}\ ,}
where 
\eqn\cBdef{C_{B\partial}= {\Gamma(2h_+)\over 2^{2h_+} 
\pi^{d/2} \Gamma(\nu)}\ .}

Finally, the boundary two-point function can be obtained by taking the
second point to the boundary,
\eqn\bdydef{G_\partial(b,b') \propto \lim_{\rho'\rightarrow\pi/2} 
(\cos\rho')^{-2h_+}
\gbdel(b,x')\ .}
This gives
\eqn\bdysum{K_\partial(b,b') = k_\partial \int {d\omega\over 2\pi} \sum_\nlm 
e^{i\omega(\tau-\tau')} {k_{nl}^2 Y_{l\vecm}^*(\ehat) Y_{l\vecm}(\ehat') \over
\onl^2-\omega^2 -i\epsilon}\ }
and 
\eqn\bdyexp{K_\partial(b,b')= C_\partial {1\over
([\cos(\tau-\tau')-\ehat\cdot\ehat' 
]^2+i\epsilon)^{h_+}}\ ,}
where $k_\partial$ and $C_\partial$ are constants.  Note that $K_\partial$
is naturally written in terms of the boundary interval \binter.

\subsec{Large-$R$ limits}

It was shown in sec.~3 that for large $R$, in a patch of proper size
$\calo(R)$, \ads\ space may be approximated by Minkowski space. This can be
explicitly seen in the coordinates 
\eqn\minkcoorda{t=R\tau\ ;\  r=R\rho\ ,}  
where the metric takes the form
\eqn\appflata{ds^2 = {1\over \cos^2(r/R)}\left[ -dt^2 + dr^2 + R^2
\sin^2({r\over R})^2 d\Omega^2\right]\ .}
This subsection will discuss other aspects of the relationship between
\mads and $M_{d+1}$ at large $R$.

First consider the relation between the symmetry generators $J_{MN}$ 
of $SO(2,d)$
and those of the Poincare group.  In the vicinity of $(\tau,\rho)=(0,0)$,
the connection can be made through the identification
\eqn\genrelat{p_{\mu}= {J_{-1\mu}\over R}\ ,\ M_{\mu\nu} = J_{\mu\nu}\ ;}
the $SO(2,d)$ algebra clearly goes over to the Poincare algebra in the
limit $R\rightarrow\infty$.  Also useful is the quadratic Casimir, which is
important for the classification of the representations of the conformal group.
It is given by
\eqn\quadc{C_2=\hf J_{MN}J^{MN} = \hf M_{\mu\nu}M^{\mu\nu} - R^2 P_\mu
P^\mu}
and takes value $C_2=-\Delta(d-\Delta)$ in a representation of conformal weight
$\Delta$.  From the relationship $\Delta=2h_+$ and eq.~\hpmdefa, we see that 
\eqn\adsmasss{C_2=m^2 R^2\ ,}
which combined with \quadc\ gives the correct mass-shell relation in
the large-$R$ limit.  The quadratic Casimir may be used to find
analogues of the Mandelstam invariants.

Next consider wavefunctions.  
The relations \enerrel, \normfreq\ imply that 
\eqn\largen{ER= 2h_+ + 2n +l\ ,}
so for fixed Minkowski energy and angular momentum, and 
$m\roughly< \calo(1/R)$, large $R$
corresponds to large $n$. 
A useful relation for large order Jacobi
polynomials is
\eqn\jacobasy{\lim_{n\rightarrow\infty} {1\over n^\alpha}
P_n^{\alpha\beta}\left(\cos(x/n)\right) = \left({2\over x}\right)^\alpha
J_\alpha(x)\ .}
For $r\ll R$ this gives
\eqn\limrad{\eqalign{ \lim_{R\rightarrow\infty}& \cos^{2h_+}(r/R) \sin^l(r/R)
P_n^{(l+d/2-1,\nu)}\left(\cos(2r/R)\right)\cr &= (ER)^{d/2-1} {1\over
(Er)^{d/2-1}} J_{l+d/2-1}(Er)\ ,}}
which is, up to the overall power of $ER$, the standard flat space radial
wavefunction.  We also need $\Anl$, which from \anldef\ via Sterling's
approximation is
\eqn\anllim{\Anl\limR \sqrt{2\onl\over R^{d-1}}\ .}
Combining this with \limrad\ then gives the desired 
wavefunctions at large-$R$:
\eqn\Rwave{\phi_{\nlm}(\vecx)\limR \sqrt{ E^{d-1}} {1\over (Er)^{d/2-1}}
J_{l+d/2-1}(Er) Y_{l\vecm}(\ehat)\ ,}
where $r\ll R $ is understood.  At $\rho\rightarrow\pi/2$, corresponding to
large $r$, one still finds behavior $\propto (\cos\rho)^{2h_+}$, and in
fact the large-$R$ limit of the coefficients $k_{nl}$ of eq.~\normasyma\ is 
\eqn\klim{k_{nl}\limR {(-1)^n \over \Gamma(\nu+1)} {\sqrt{2E}\over R^{d/2-1}}
 \left({ER\over
2}\right)^\nu\ .}

Finally consider the large-$R$ behavior of the Green functions.  The
asymptotics of the bulk propagator immediately follows from \feyngreen.  Geodesic
distance on \mads\ trivially becomes the Minkowski interval.  The
hypergeometric function must therefore 
be evaluated with argument near one, which is
done via the formula
\eqn\hypergeqn{\eqalign{F(h_+&,h_++\hf;\nu +1; z) = {\Gamma( \nu+
1) \Gamma({1-d\over 2}) \over \Gamma(1-h_-) \Gamma(\hf-h_-) } 
F(h_+,h_++\hf;{d+1 \over 2} ;1- z)\cr & + (1-z)^{(1-d)/2} {\Gamma( \nu+1) 
\Gamma({d-1\over 2}) \over \Gamma(h_+ ) \Gamma(h_++\hf) }
F(1-h_-,\hf-h_-;{3-d\over 2};1- z)\ .}}
Taking
\eqn\zlimit{z={1\over \cosh^2(s/R)} \approx 1-{s^2\over R^2}\ ,}
and assuming that $\nu$ stays finite as $R\rightarrow\infty$, we find 
\eqn\gBlim{iK_B(x,x')\limR  {{\tilde C}\over [s^2(x,x')+
i\epsilon]^{(d-1)/2}}\
,}
where 
\eqn\tilC{{\tilde C} = { \Gamma({d-1\over 2}) \Gamma(2h_+)\over \pi^{d/2}
2^{2h_+ +1} \Gamma(h_+) \Gamma(h_+ +\hf) }\ .}
This is the 
expected massless flat-space propagator.

As we saw in sec.~4, for generic frequencies the bulk-boundary
propagator $\kbdel$ grows like $(\cos\rho)^{2h_-}$ at the boundary,
and so in the large-$R$ limit is not concentrated in the Minkowski
region.  This can be remedied by restricting to normalizable
frequencies, as in eq.~\newbb.  The large-$R$ behavior of the
resulting function ${\hat K}$ is readily inferred from \Rwave\ and
\klim, and gives
\eqn\newlima{\khat(n,l,\vecm,x)\limR C(E,R) (-1)^{l/2} 
{1\over (Er)^{d/2-1}} J_{l+d/2-1}(Er) Y_{l\vecm}^* (\ehat)\ }
where the coefficient function $C(E,R)$ is given by
\eqn\Cerdef{C(E,R)={2^{2-\nu}\over \Gamma(\nu) } (-1)^{ER/2-h_+}
(ER)^{2h_+}\ .}

\listrefs
\end